\documentstyle[dina4,12pt,psfig]{article}
\begin{document}
\title{Antibaryons in massive heavy ion reactions:
  Importance of potentials}
\author{C.~Spieles, M.~Bleicher, A.~Jahns, R.~Mattiello,\\ H.~Sorge, H.
St\"ocker, W.~Greiner\\
{\it Institut f\"ur Theoretische Physik, J.~W.~Goethe--Universit\"at,}\\
{\it 60054 Frankfurt am Main, Germany} }

\maketitle

\begin{center}
Intended for: Relativistic/Ultrarelativistic Nuclear
Collisions, Brief Reports
\end{center}

\begin{abstract}
In the framework of
RQMD we investigate antiproton observables in massive heavy
ion collisions at AGS energies and compare to preliminary results of the
E878 collaboration. We focus here on the considerable influence
of the {\it real} part of an antinucleon--nucleus optical potential
on the $\bar p$ momentum spectra.\\[0.01cm]
\end{abstract}


\noindent Antibaryon production is a promising observable for collective
effects in nucleus--nucleus collisions. On the other hand, $\overline N$'s
suffer strong final--state interactions. These interactions have two
components
which can be related to the $\overline N$ self--energy in matter: collisions
and annihilation on baryons\cite{Gav90} (imaginary part,
semiclassically given by $ 2{\rm Im}V=\sigma v \rho $)
and a piece in the real part
($ {\rm Re}V=t_{\rm NN}\rho$ in the impuls approximation).
 In the semiclassical limit the real part of the self-energy
 can be approximated by
potential--type interaction\cite{Dover} or a mean field.

Here we will focus on the effect of the real part. The motivation is that
the long--range force of baryons acting on a $\bar p$ is expected to be
stronger than for protons since the Lorentz--scalar and the Lorentz--vector
parts of a meson exchange potential now have the same sign.
The influence of baryonic mean--fields on baryons and mesons
is well established. Therefore there should also be some influence on $\bar
p$'s.

\noindent The substantial impact of mean--fields on particle spectra was
studied earlier \cite{VKoch,Shu91}. Following these ideas, we now investigate
observables in nucleus--nucleus collisions,
 where $\overline B B$-potentials come into play. For this purpose, we
employed a simple model--interaction, knowing that this choice is far
from being unique.

\begin{figure}[p]
\centerline{\psfig{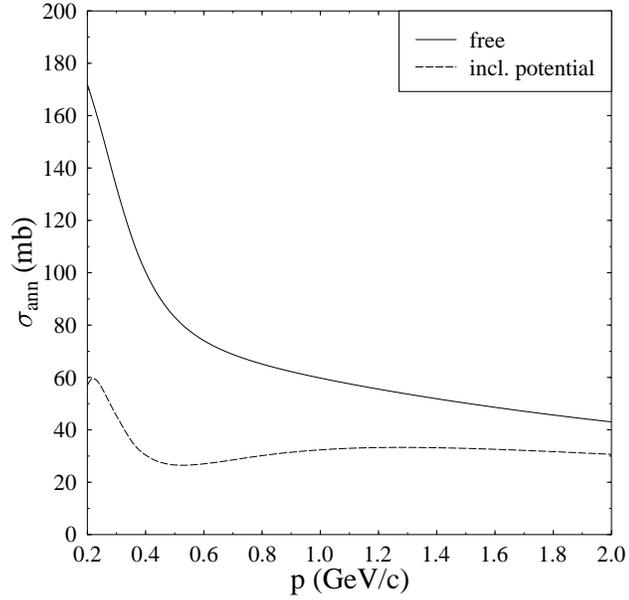}}
\vspace*{-1cm}
\caption[]{$\overline p p$ annihilation cross section as a function of $p_{\rm
lab}$. Parametrisation of the free measured cross section (full line) and
the corrected value, if potential interaction is added (dashed line).
\label{RANNTRUE}}
\end{figure}

\noindent
In principle,
one has to calculate the medium--dependent mean--field and the cross--section
selfconsistently to understand $\overline N$ behaviour in matter. We
calculate the forces acting on a $\overline N$ in a baryonic medium only
{\em after} freeze--out. However, by taking the free interaction ---
annihilation, elastic scattering, non--annihilating inelastic channels ---
for $\overline NN$ in the collision term during the dynamical evolution
the real potential is included effectively.
Our approach is similar in its spirit   to
the usual treatment of the Coulomb distortion
of particle spectra which is also
restricted to final--state interactions\cite{Gyu81}.
Addition of free interactions and mean--field
contribution would cause  double counting of interactions: any
geometric cross section will appear as a larger physical cross section if
the particles' trajectories are bent due to attractive forces.
Fig.~\ref{RANNTRUE} shows this effect
for our model potential. Due
to the strong attraction for the $\bar B B$ case, a reduced geometric
annihilation cross section suffices to account for the measured free
annihilation probability in binary $\bar N N$ reactions.
This is in rough correspondence to phenomenological models of $\bar p p$
interaction \cite{Weise}, where the characteristic range of the imaginary
potential is chosen to be less than 1~fm, which corresponds to a disc
much smaller than the free annihilation cross section.
 For the same reason the insertion of {\em free} cross sections in
the collision term of Vlasov--type models seems questionable.

\noindent The great success of Dirac equation optical model calculations for
$pA$
scattering \cite{Arn81} led us to using
 these relativistic
potentials  (Lorentz--scalar and
 vector interaction) with Yukawa functions as
 form factors
 --- applying G--parity transformation --- for the $\overline
p$ case
\footnote{Optical model calculations for intermediate energy antiproton
scattering like in \cite{Kubo85,Ing86} cannot provide unambigous values
of the real part of the optical potential, since the imaginary part
dominates the  behavour.}:
The mass paramters are $\mu_V=3.952\;\rm fm^{-1}$ and
$\mu_S=2.787\;\rm fm^{-1}$, the coupling constants are
$g_V=13.5\;\rm MeV\,fm$ and $g_S=10.9\;\rm MeV\,fm$.
In line
with \cite{Arn81} we used gaussians as baryon profiles with a
mean square radius of $0.8\;\rm fm$.
The central part of
an effective Schr\"odinger equivalent potential (SEP) is constructed from the
above potentials:
\[
U_{\rm SE}=\frac{1}{2E}(2EU_V+2mU_S-U_V^2+U_S^2),
\]
where $E$ is the total energy of the incident particle.

The $\overline p$'s are produced and propagated with the usual RQMD
simulation \cite{RQMD1} ($NN$-potentials are switched off, i.~e.
pure cascade mode) until they
undergo the last strong interaction. Afterwards the trajectories of the
$\bar p$'s are calculated
under the influence of  the relativistic
optical potentials, while the remaining particles continue to interact
 within the cascade calculation.

\noindent For $\overline N N$ distances,
smaller than the corresponding annihilation
cross section,
the real potential should not show an effect, since the huge
imaginary part absorbs the particular $\overline p$ anyhow. We have chosen
1.5~fm
as cutoff distance
 corresponding
 to an averaged $\overline p p$-annihilation cross section
of $\approx 70$~mb.

\noindent The Schr\"odinger equivalent potential with the above parameters
results in
a mean $\overline p$-potential in nuclear matter
at groundstate density of about $-250\;\rm MeV$
($p_{\overline p}=0\;\rm GeV/c$), increasing with energy towards $-170\;\rm
MeV$
at $p_{\overline p}=1\;\rm GeV/c$.
These values are much more in accord with
estimations on the basis of dispersion relations \cite{Teis}
than just taking the sum of
the scalar and vector potentials without any cutoff.
The actual (averaged) SE-potential of the $\overline p$'s
at freeze--out is about $-70$~MeV in central
$Au+Au$ collisions at 10.7~GeV. We have also studied the influence
of Coulomb--effects on the $\bar p$ momentum spectrum. It shows
that the average potential is less than $-10$~MeV and therefore of less
importance
\footnote{The effect may be
relevant for pions: in the above system the ratio $\pi^-/\pi^+$ for
$p_t<100$~MeV increases according to our simulation
by 25 percent due to the Coulomb potential.}.

\noindent
For central collisions of $Au+Au$ at 10.7~GeV we
find a substantial change in the final
phase space distribution of the $\overline p$'s at low--$p_t$, deviating from
results of a pure cascade, due to the potential interaction
during the last stage of the collision.
Without potentials, we find
a clearly nonthermal spectrum with a dip at
midrapidity for $p_t=0$.
Due to the strong momentum dependence of the $\overline p
p$-an\-ni\-hi\-la\-tion cross
section, $\overline p$'s with
low transverse momentum are suppressed.

\begin{figure}[p]
\centerline{\psfig{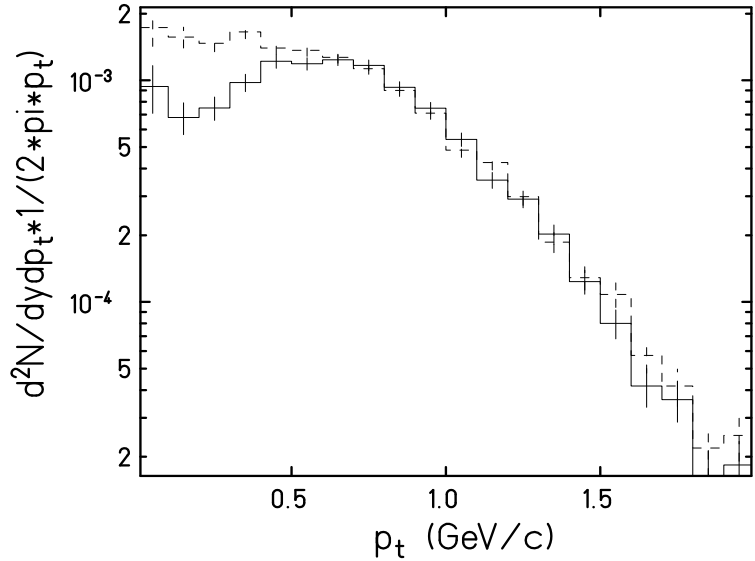}}
\vspace*{-0.25cm}
\caption[]{Invariant $p_t$--spectrum of the $\overline p$'s at midrapidity
($-0.4<y<0.4$) for central reactions of $Au+Au$ ($b<4$~fm) at 10.7~GeV/u.
Shown is the spectrum of the RQMD calculation (full line) and with an
additional optical potential (dashed line).
\label{PT}}
\end{figure}

\begin{figure}[p]
\centerline{\psfig{figure=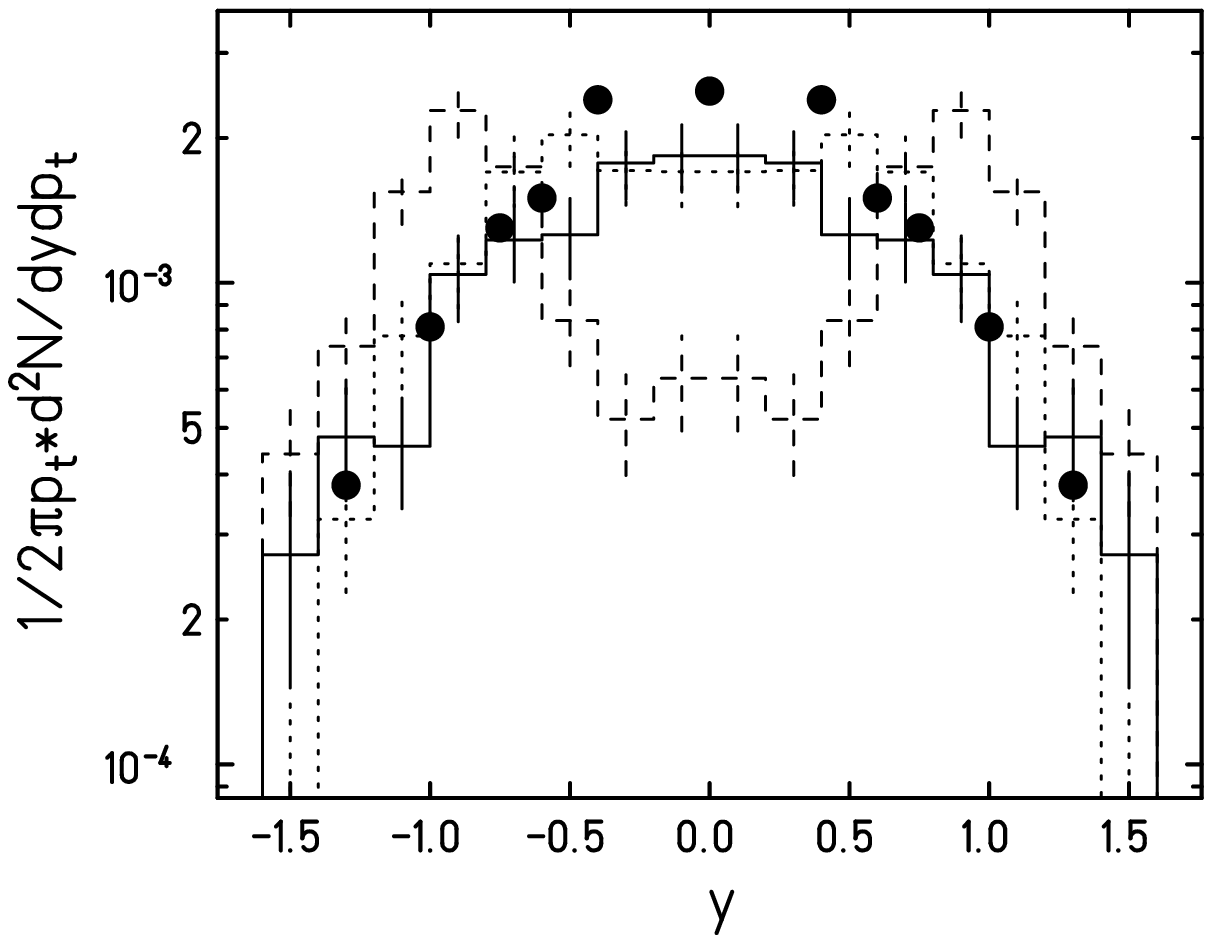,width=10cm,height=7.5cm}}
\vspace*{-0.25cm}
\caption[]{
Invariant rapidity--distribution of the $\overline p$'s
with $p_t<200$~MeV for $Au+Au$ ($b<4$~fm) at 10.7~GeV/u.
Shown is the RQMD calculation (dashed line), with the
additional optical potential (full line) and the 50~\% reduced potential
(points). Preliminary data
(full circles) from \cite{E878} (the error bars are omitted).
\label{DNDY}}
\end{figure}

Figure~\ref{PT} shows the $p_t$-spectrum at midrapidity
($-0.4<y<0.4$)
of the antiprotons
with and without
the inclusion of potential interaction.
With potential interaction included
the dip at low $p_t$ gets filled without changing the shape of the
spectrum at higher momenta.
Figure~\ref{DNDY} shows the invariant multiplicity $\displaystyle
\frac{1}{2\pi
p_t}\frac{{\rm d}^2N}{{\rm d}y{\rm d}p_t}$ of $\overline p$'s with
$p_t<200\;\rm MeV$. RQMD 1.07 calculations with and without potentials are
compared to preliminary data of the E878 collaboration\cite{E878}.
in addition to the presented result for
the proposed model interaction we calculated the effect for the same
potentials
arbitrarily reduced by 50~\%. Still the dip at midrapidity vanishes although
the change is less pronounced.
Note that the $p_t$-integrated spectrum
at midrapidity is affected by less than 25 percent
in comparison to the calculation  without potential
interaction.

$\left. \right.$

The theoretical understanding of $\bar p$ propagation in a baryon--dense
medium is far from satisfying. For instance, the absorption strength differs
by orders of magnitude if one compares different transport
calculations$[11,13-18]$.
Our study should be seen as   a  first step on
the way towards a self-consistent treatment of the real and the imaginary
parts of the $\overline NN$ potential in a
transport theoretical calculation.



\begin{thebibliography}{99}
%
\bibitem{Gav90} S.~Gavin, M.~Gyulassy, M.~Pl\"umer, R.~Venugopalan, {\em Phys.
Lett. {\bf B 234} (1990) 175}
%
\bibitem{Dover}
W.~W.~Buck, C.~B.~Dover, J.~M.~Richard, {\em Ann. Phys. (N.Y.) {\bf 121}, 47
(1979)}; C.~B.~Dover, J.~M.~Richard, {\em ibid., p.~70}
%
\bibitem{VKoch} V.~Koch, G.~E.~Brown, C.~M.~Ko, {\em Phys. Lett. {\bf B
265} (1991) 29}
%
\bibitem{Shu91} E.~V.~Shuryak, {\em Nucl. Phys. {\bf A533} (1991) 761}
%
\bibitem{Gyu81} M.~Gyulassy and S.~K.~Kauffmann, {\em Nucl. Phys. {\bf A362}
(1981) 503}
%
\bibitem{Weise} M.~Kohno, W.~Weise: {\em Nucl. Phys. {\bf A454} (1986) 429}
%
\bibitem{Arn81} L.~G.~Arnold, B.~C.~Clark, R.~L.~Mercer, P.~Schwandt,
{\em Phys. Rev. {\bf C 23} (1981) 1949}
%
\bibitem{Kubo85} K.--I.~Kubo, H.~Toki, M.~Igarashi {\em Nucl. Phys. {\bf
A435} (1985) 708}
%
\bibitem{Ing86} A.~Ingemarsson {\em Nucl. Phys. {\bf A454} (1986) 475}
%
\bibitem{RQMD1} H. Sorge, H. St\"ocker, W. Greiner, {\em Ann. Phys. (N.Y.)
{\bf 192} (1989)
266}
%
%
%
\bibitem{Teis} S.~Teis, W.~Cassing, T.~Maruyama, U.~Mosel, {\em Phys. Rev.
{\bf C 50} (1994) 388}
%
\bibitem{E878} M.~J.~Bennett for the E878 collab., proceedings of the Quark
Matter
'95, to be published in {\em Nucl. Phys. {\bf A}}
%
\bibitem{RVUU} G.~Q.~Li, C.~M.~Ko, X.~S.~Fang, Y.~M.~Zheng,
{\em Phys. Rev. {\bf C 49} (1994) 1139}
%
\bibitem{Faessler} G.~Batko, A.~Faessler, S.~W.~Huang, E.~Lehmann,
R.~K.~Puri, {\em J.~Phys.~{\bf G 20} (1994) 461}
%
\bibitem{ARC} S.~H.~Kahana, Y.~Pang, T.~Schlagel, C.~B.~Dover,
{\em Phys.~Rev.~{\bf C 47}, R1356 (1993)}
%
\bibitem{CSP}
C. Spieles, A. Jahns, H. Sorge, H. St\"ocker, W. Greiner,
{\em Mod. Phys. Lett {\bf A 27}, (1993) 2547}
%
\bibitem{pbar1}
A. Jahns, C. Spieles, R. Mattiello, H. St\"ocker, W.Greiner, H. Sorge,
{\em Phys. Lett {\bf B 308}, (1993) 11}; A.~Jahns, C.~Spieles, H.~Sorge,
H.~St\"ocker, W.~Greiner,
      {\em Phys. Rev. Lett. {\bf 72} (1994) 3464}
%
\bibitem{pbar3}
H. Sorge, M. Berenguer, H. St\"ocker, W. Greiner,
{\em Phys. Lett. {\bf B 289} (1992) 6}
\end{thebibliography}
\end{document}